\documentclass[aps,prd,amsmath,twocolumn,preprintnumbers]{revtex4-1}
\usepackage{amssymb,esvect,amsmath,graphicx,latexsym,amsthm,slashed,eso-pic,hyperref}
  \DeclareMathOperator{\mev}{MeV} \DeclareMathOperator{\gev}{GeV}  \DeclareMathOperator{\cm}{cm} \DeclareMathOperator{\barn}{barn} \DeclareMathOperator{\g}{g} \DeclareMathOperator{\km}{km}  \DeclareMathOperator{\s}{s}    \DeclareMathOperator{\few}{few} 
 \newcommand{\cB}{{\cal B}}      \newcommand{\cM}{{\cal M}}  \newcommand{\cO}{{\cal O}}

\newcommand{\vAcm}{v_{1, {\rm cm}}}

\newcommand{\sff}{\sigma_{\rm f}}
\newcommand{\sel}{\sigma_{\rm el}}

  \newcommand{\eg}{{\it e.g.~}}
   \def\oL{\overline}
\newcommand{\pL}{\left(} \newcommand{\pR}{\right)} \newcommand{\bL}{\left[} \newcommand{\bR}{\right]} \newcommand{\cbL}{\left\{}  \newcommand{\mL}{\left|} \newcommand{\mR}{\right|}
\newcommand{\beq}{\begin{equation}} \newcommand{\eeq}{\end{equation}}
\newcommand{\bea}{\begin{eqnarray}} \newcommand{\eea}{\end{eqnarray}}
\newcommand{\alg}[1]{\begin{align} \begin{split} #1 \end{split}  \end{align}}

\newcommand{\tenx}[1]{\times 10^{#1}}
\newcommand{\Eq}[1]{Eq.~(\ref{#1})} \newcommand{\Eqs}[2]{Eqs.~(\ref{#1}) and (\ref{#2})} 
\newcommand{\Sec}[1]{Sec.~\ref{#1}}  
\newcommand{\Fig}[1]{Fig.~\ref{#1}}

\begin{document}

\title{Is Self-Interacting Dark Matter Undergoing Dark Fusion?}
\author{Samuel D. McDermott}
\affiliation{Fermi National Accelerator Laboratory, Center for Particle Astrophysics, Batavia, IL, USA}
\date{\today}

\begin{abstract}
We suggest that two-to-two dark matter fusion may be the relaxation process that resolves the small-scale structure problems of the cold collisionless dark matter paradigm. In order for the fusion cross section to scale correctly across many decades of astrophysical masses from dwarf galaxies to galaxy clusters, we require the fractional binding energy released to be greater than $v^n \sim [10^{-(2-3)}]^n$, where $n=1,2$ depends on local dark sector chemistry. The size of the dark-sector interaction cross sections must be $\sigma \sim 0.1-1\barn$, moderately larger than for Standard Model deuteron fusion, indicating a dark nuclear scale $\Lambda \sim \cO(100\mev)$. Dark fusion firmly predicts constant $\sigma v$ below the characteristic velocities of galaxy clusters. Observations of the inner structure of galaxy groups with velocity dispersion of several hundred kilometer per second, of which a handful have been identified, could differentiate dark fusion from a dark photon model.
\end{abstract}
\preprint{FERMILAB-PUB-17-483-A-T}

\maketitle

\section{Introduction}

A dark, stable form of matter is overwhelmingly successful at explaining astrophysical phenomena on the largest observable scales of time and space. Nevertheless, the inner structure of some dark matter potential wells may indicate discrepancies with expectations derived from the simplest models of dark matter \cite{Bullock:2017xww, Tulin:2017ara}. Whether these discrepancies are due to mismodeled but entirely standard physics or exotic, ``beyond the Standard Model'' particles and forces is one of the central questions for the present generation of physicists. It has recently been noted that dark matter that interacts with itself with a nuclear-scale cross section can compactly account for many of these observations \cite{Kaplinghat:2015aga, Kamada:2016euw}. Intriguingly, the required scattering cross section times velocity seems to be constant for objects across several decades of velocity dispersion.

In parallel, investigations of the early-universe cosmology of models with dark-sector forces have led to some striking conclusions. A number of authors working with a wide variety of model assumptions have concluded that synthesis of bound states of dark matter particles is a generic consequence of the high temperature history of such hidden sectors \cite{Krnjaic:2014xza, Detmold:2014qqa, Wise:2014jva, Wise:2014ola, Hardy:2014mqa, Forestell:2016qhc, Gresham:2017cvl, Gresham:2017zqi, Forestell:2017wov}. This implies that dark matter particles with a range of ``dark nucleon number'' could be present and fusing in dark matter halos today.

In this paper, we link these developments by demonstrating that late-time two-to-two fusion of dark matter particles has a velocity-dependent cross section that can resolve small-scale structure ``crises'' from the scales of dwarf galaxies to galaxy clusters. The simple relations for two-to-two fusion discussed here provide a firm prediction of an {\it exactly flat} $\sigma v$ at intermediate and low velocities. This contrasts with the model presented in \cite{Kaplinghat:2015aga, Tulin:2012wi, Tulin:2013teo}, which exhibits a peak in $\sigma v$ at velocities between galaxy and cluster scales. The reason fusion is a powerful explanatory mechanism is that dwarf galaxies probe much smaller velocities than galaxy clusters, but the relaxation rates of these halos seem to be similar \cite{Kaplinghat:2015aga, Kamada:2016euw}. It may be possible to distinguish these models by detailed observation of galaxy groups, which are virialized and have velocity dispersion in an interesting intermediate range.

This model is not in jeopardy from limits on unitarity of scattering \cite{Hui:2001wy} because the particles are relatively light and can be compound objects, and it evades observational limits on the Standard Model flux from products of annihilation \cite{Kaplinghat:2000vt, Beacom:2006tt} if the initial states do not annihilate or if the final states are low energy or stable against decay to Standard Model particles. In this sense, dark fusion is a ``gentle'' generalization of annihilation in which much less rest mass energy is released and antiparticles need not be present. Decoupling the annihilation and energy release mechanisms opens up a bevy of new model-building opportunities and naturally allows a large cross section with less severe consequences. The large cross sections discussed here are also not problematic for direct detection since they connect the dark sector to itself, and the dark sector may be arbitrarily secluded from the Standard Model. Dark fusion is thus able to explain a wide variety of observations, while evading nontrivial bounds, in a compelling way.

In \Sec{sec:vel-dep} we will discuss how the required velocity dependence arises. We show that there are two distinct limits of the annihilation participant masses that can solve this problem, and in \Sec{sec:results} we discuss the phenomenological requirements on (and benefits of) such models. In \Sec{sec:models} we provide an outline of some complete models whose early universe dynamics could satisfy these phenomenological requirements. Finally, we conclude.

\section{Two-to-Two Scattering and Cross Section Velocity Dependence}
\label{sec:vel-dep}

The scattering cross section for a two-to-two process is
\beq \label{dsig-def}
\frac{d\sigma}{d\Omega} = \frac{\oL{\mL \cM \mR}^2}{64\pi^2s} \frac{\mL \vec p_3\mR}{\mL \vec p_1\mR},
\eeq
where the momenta are in the center of mass frame, with $\mL \vec p_1\mR = \mL \vec p_2\mR$ and $\mL \vec p_3\mR = \mL \vec p_4\mR$. These momenta may be expressed solely in terms of $s$, the center of mass energy squared, and the particle masses:
\beq \label{momentum}
\mL \vec p_1\mR= \frac{\sqrt{\bL s-(m_1+m_2)^2 \bR \bL s-(m_1-m_2)^2  \bR}}{2\sqrt s}.
\eeq
For $\mL \vec p_3\mR$ we take $m_i \to m_{i+2}$ in \Eq{momentum}. Without loss of generality we will choose $m_1 > m_2$ and $m_3 > m_4$ if the masses are unequal.

We focus on exothermic fusion, $m_3+m_4 < m_1+m_2$, and define the fractional binding energy released as
\beq
b=1-\frac{m_3+m_4}{m_1+m_2}.
\eeq
Kinematically unsuppressed annihilation is the limit $b\sim 1$, and semiannihilation \cite{Boehm:2014bia, Ko:2014gha} is $b\sim 0.5$. By analogy with Standard Model nuclear forces, and for observational reasons discussed in more detail below, we will be interested in the regime $b\ll1$. Furthermore, we will assume that the early-universe yields of dark nuclei lead to a few dominant dark species with similar dark nucleon number, which implies that the initial state masses are comparable. A UV complete model simultaneously displaying all necessary ingredients has not been explicitly demonstrated to exist, but in \Sec{sec:models}, we discuss models that plausibly have both $0<b\ll 1$ and appropriately heterogeneous initial state abundances. Finally, we define a velocity via $E_{i{\rm,cm}} = m_i(1-v_{i{\rm,cm}}^2)^{-1/2}$, so that in the nonrelativistic limit $ \mL \vec p_1\mR = m_1\vAcm $.

For now, consider the scenario where $b\ll1$ and the outgoing dark nucleons have similar nucleon numbers, allowing us to write $s- (m_3-m_4)^2 \simeq s- (m_1-m_2)^2 \simeq 4m_1 m_2$. In the nonrelativistic limit and to leading order in $b$, the final-state momenta are
\alg{ \label{momenta-NR}
\mL \vec p_3\mR_{m_4\simeq m_3} & \simeq  m_1 \pL \frac{2 m_2}{m_1}b +\vAcm^2 \pR^{1/2}.
}
The kinematic assumptions break down and \Eq{momenta-NR} is changed if the mass of the lighter outgoing particle is much smaller than the binding energy. Consider a model of dark nucleons, with dark nucleon mass $m_d$, dark nucleon number $A_i$, and binding energy $\cB_i$. A light outgoing particle corresponds to $A_4/A_3 \ll b$. This occurs with bremsstrahlung of a light, non-nuclear partner particle, so that $A_1 + A_2 = A_3$. A familiar example is Standard Model deuteron production, during which a photon is emitted. For this case,
\beq \label{momenta-A40}
\mL \vec p_3\mR_{\frac{m_4}{m_3} \ll b} \simeq m_1 \pL \frac{1+ m_1/m_2}2 \pR \pL\frac{2 m_2}{m_1} b+\vAcm^2 \pR.
\eeq
In either limit the flux factor $\mL \vec p_3\mR/\mL \vec p_1\mR$ goes like $ \sim \vAcm^{-1}$ at velocities below the square root of the fractional binding energy released. Thus, fusion is qualitatively different than elastic scattering, for which $\mL \vec p_3\mR = \mL \vec p_1\mR$ is exact. The underlying difference is that in the case of fusion, $\mL \vec p_3\mR$ need not be small if significant binding energy is released, even with initial state particles at rest. In this way, rates for exothermic self-interactions can be constant at low velocities. Regardless of the hierarchy of $m_4$ and $m_3$, the comparison $2 b m_2/m_1 \lessgtr \vAcm^2$ is the correct diagnostic for determining whether $|\vec p_3|$ has a velocity dependence. The dependence on velocity in the high-velocity limit is linear when $m_4/m_3 \sim \cO(1)$ and quadratic when $m_4/m_3$ is very small.

Given the discussion above, we define a function $f(b,\vAcm )\equiv \mL \vec p_3\mR/m_1$, which has the limiting forms
\alg{ \label{flux-func}
f(b,\vAcm ) = \cbL \begin{array}{cc}  \pL \frac{2 m_2}{m_1}b +\vAcm^2 \pR^{1/2} & \frac{m_4}{m_3} \gtrsim b 
\\ \frac{1+ m_1/m_2}2 \pL \frac{2 m_2}{m_1} b+ \vAcm^2 \pR & \frac{m_4}{m_3} \ll b  \end{array} \right.
}
Using \Eqs{dsig-def}{flux-func}, we may then write the nonrelativistic fusion cross section simply as
\beq \label{svcm}
\sff \vAcm = f(b,\vAcm) \bar \sigma (m_1, m_2),
\eeq
where we define $\bar \sigma (m_1, m_2) \simeq \frac{1}{16\pi (m_1 + m_2)^2} \int \frac{d\Omega}{4\pi} \oL{\mL \cM \mR}_{\rm f}^2$. For non-relativistic two-to-two scattering with a constant matrix element, $\sff \vAcm$ inherits velocity dependence only from the flux factor $f(b,\vAcm)$.

For these reasons, it is necessary to understand the velocity dependence of $\bar\sigma$. In \cite{Krnjaic:2014xza}, the fusion cross section is estimated as $\sigma_u =  \pL \frac{A_1^{1/3}+A_2^{1/3}}{\Lambda_u} \pR ^2 \exp\pL - \frac{\alpha_d A_1A_2 }{\vAcm/\sqrt2} \pR$ in a dark nuclear model described in more detail in \Sec{sec:models}. The exponential term describes the action for tunneling through the Coulomb barrier. At low velocities the Coulomb factor will saturate, the flux term will become nonnegligible, and this prescription for the cross section must break down. By analogy with nuclear physics, the cross section is still expected to scale like the dark nucleon number to the 2/3 power, \eg like the particle radius squared assuming a constant dark nuclear density \cite{Hardy:2014mqa}. Knowledge of nonperturbative dark sector physics is required to perform a first-principles calculation of the precise manner in which the cross section saturates to this geometric limit \cite{Gresham:2017cvl, Gresham:2017zqi}. However, the magnitude is parametrically set by a dark nucleon scale $\Lambda_d$. Assuming couplings $\sim \cO(1)$, we will speculatively write $\bar \sigma \equiv 1 / \Lambda_d^2$. For a realistic model, it is natural to assume that elastic scattering that does not change dark nucleon number is {\it also} generically present with a similar size, $\sel \sim \bar \sigma$. We include elastic processes in what follows. The scale $\Lambda_d$ will be set by the requirement that $\sff \vAcm$ matches observations, and will turn out to be explained with $\Lambda_d \sim m \sim \cO(100\mev)$.

\begin{figure*}[t]
\begin{center}
\includegraphics[width=0.48\textwidth]{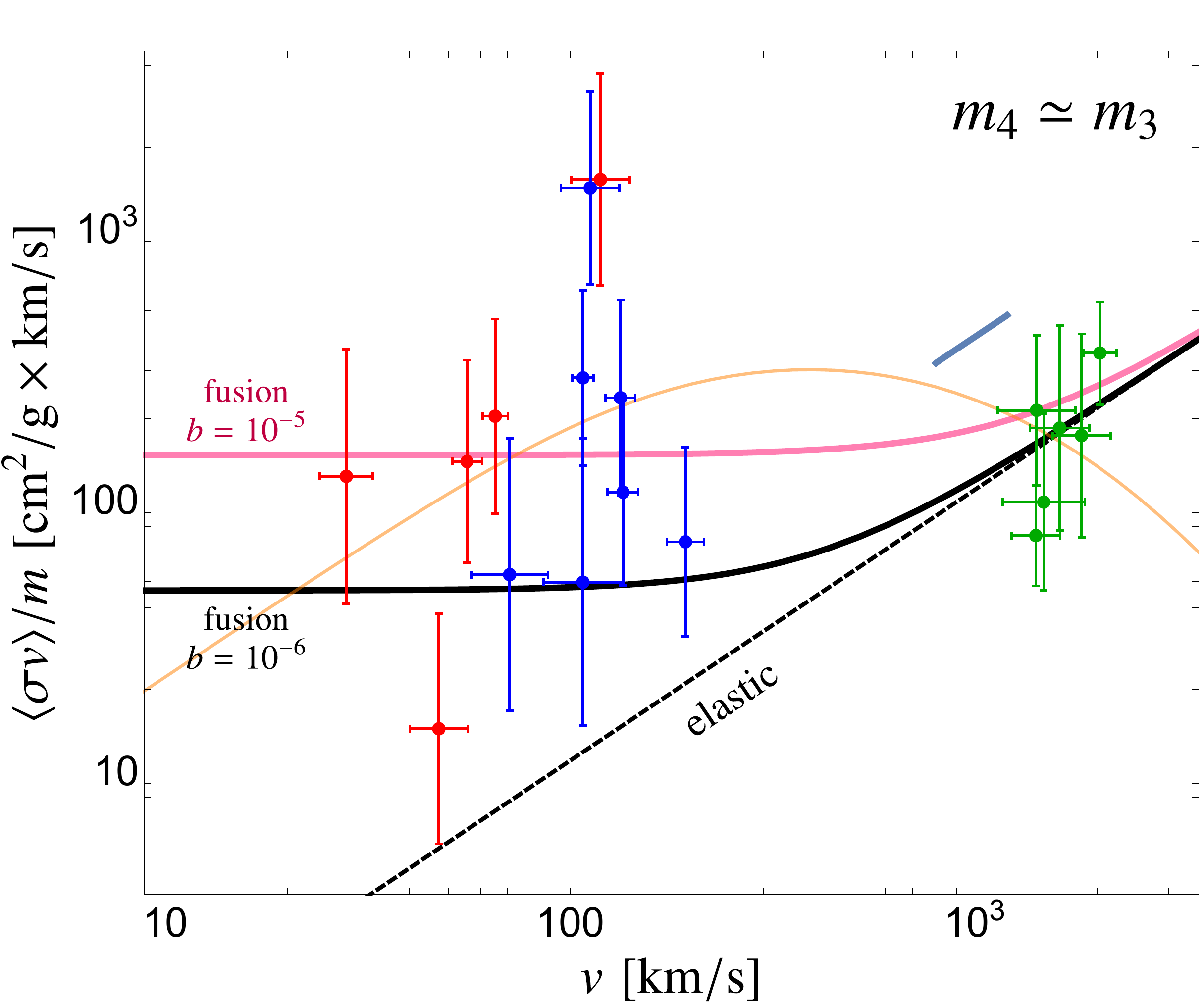}~~~
\includegraphics[width=0.48\textwidth]{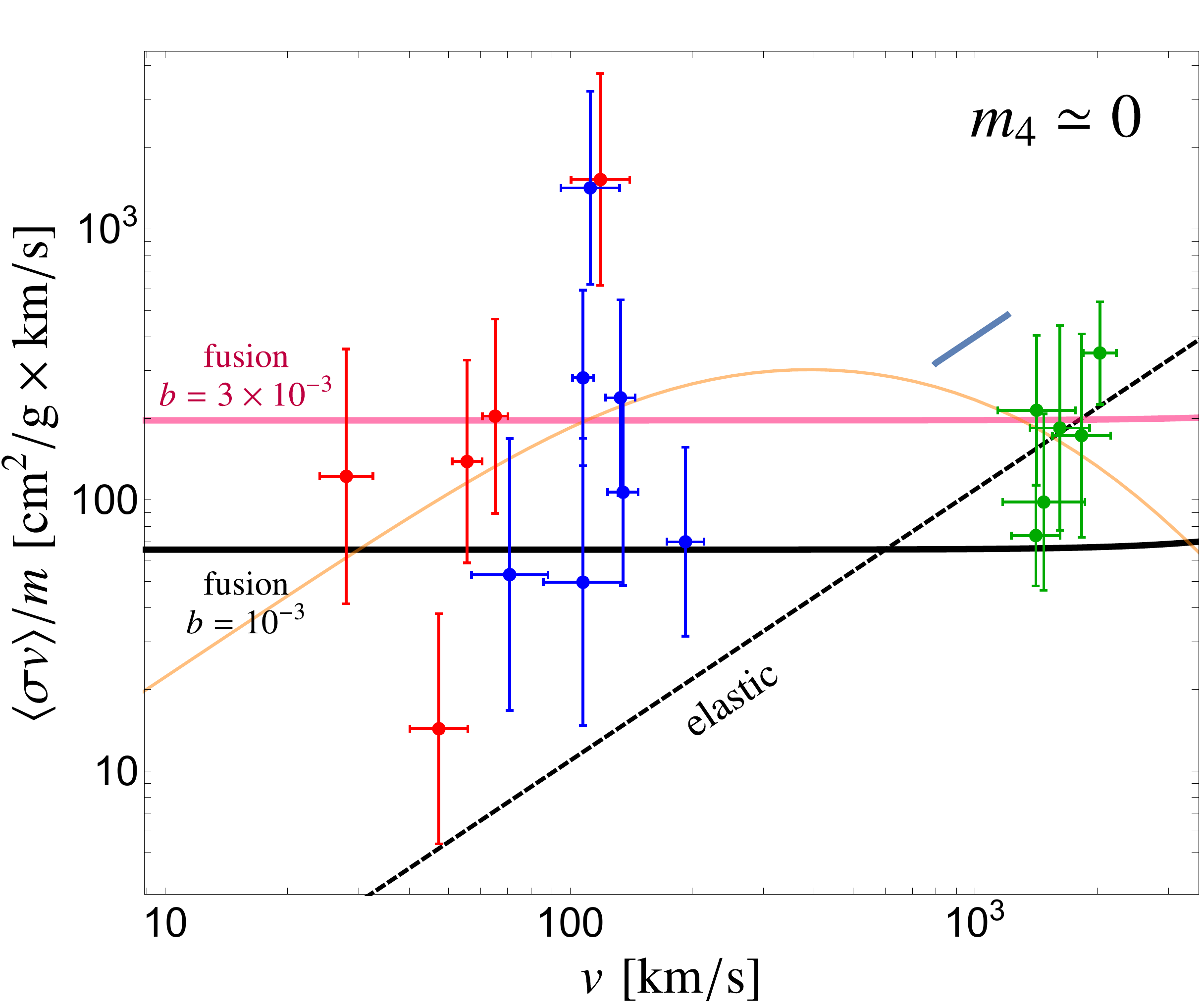}
\caption{The dark fusion (thick solid) and elastic (thick dashed) scattering cross sections. We assume $m_1 = m_2 = 200\mev$, we fix the elastic scattering cross section $ \sel =  (100\mev)^{-2}$, and we set $\sff v = f(b,v) \sel$. In the {\bf left} panel we assume that all reactants have similar mass. In the {\bf right} panel we assume that one outgoing fusion product is massless. Points with error bars provide locations of interesting anomalies in the inner structure of dark matter halos and the thin orange line is a best fit dark photon model to this data \cite{Kaplinghat:2015aga}. The short, thick blue line in the upper right corner of each plot shows a rough bound of $\bar \sigma \leq 0.4\cm^2 \!/\g$ from stacked observations of cluster mergers at $v= 1000 \pm 200\km/\s$ \cite{Harvey:2015hha}.}
\label{plot}
\end{center}
\end{figure*}

Collecting these observations, the velocity dependence of $\sff \vAcm$ has two distinct regimes: at high velocities $\sff \vAcm \sim \vAcm^{1,2}$, where the exponent depends on the mass ratio $m_4/m_3$, while {\it at low velocities there is no velocity dependence}, regardless of the value of $m_4/m_3$. Early-universe nucleosynthesis occurs at temperatures corresponding to velocities $v^2 \sim \cO(1/20)$, for which the high-velocity limit is suitable. The physical intuition is that if the mass difference is small compared to the kinetic energy the scattering is nearly elastic. On the other hand, inelasticity is crucial at low velocity. We see from \Eq{flux-func} that the {\it product} $\sff \vAcm$ asymptotes to be constant with velocity, $\sff \vAcm|_{{\rm low\,}v} \propto v^0$. Because dark matter halos provide potential wells with virial velocities $v \sim \sqrt{GM/R} \sim 10^{-(3\pm1)}$, the case $\sff \vAcm \propto \vAcm^0$ is appropriate for virialized particles in galaxies at late times as long as $b$ is not arbitrarily small.

\section{Dark Fusion at Late Times}
\label{sec:results}

Some decades ago, Bethe and Longmire \cite{Bethe:1950jm} showed that the low-velocity cross section for production of a deuteron along with bremsstrahlung of a photon is \footnote{We thank Susan Gardner for emphasizing the importance and generality of this result.}
\beq \label{B-L}
\sigma_{\rm H} v \simeq 6.64 \tenx{-20} \cm^3/\s.
\eeq
The size of this cross section is set by the mass scale of the nucleons $\sim 1 \gev$ and the fractional binding energy of deuterium, $1-\frac{m_D}{m_n+m_p} \sim 10^{-3}$. More recently, the preferred cross section for explaining the cores of dark matter halos through self-interacting dark matter has been claimed to be \cite{Kaplinghat:2015aga, Kamada:2016euw}
\beq \label{SIDM-favored}
\frac{\langle \sigma_s v \rangle}{m_X} \simeq  \frac{40-200 \cm^2}\g \frac \km \s  \simeq \frac{6-30\tenx{-18} \cm^3}{\gev\s}.
\eeq
Particles with this scattering cross section will interact roughly once in a Hubble time, thermalizing a core in their host halo \cite{Kaplinghat:2015aga}. The cross section in \Eq{SIDM-favored} is suggestively close to the range in \Eq{B-L}, albeit somewhat larger. Critically, both \Eqs{B-L}{SIDM-favored} are much larger than the thermal annihilation cross section. This is understandable if the early universe abundances are set by an asymmetry rather than weak self-annihilation -- but such an asymmetry would render the current population non-annihilating. This both resolves the problem and removes the solution of extreme annihilations at the center of dark matter halos \cite{Kaplinghat:2000vt}. Instead, we speculate that SIDM scattering is dark fusion mediated by a dark sector interaction with a scale slightly below the nuclear scale.

We show the fusion and elastic cross sections times velocity $\sff \vAcm$ and $\sel \vAcm$ in \Fig{plot}. We relate the fusion to the elastic scattering cross section by the flux factor, $\sff \vAcm = f(b,\vAcm) \sel$ under the assumption that all dark sector dynamics are at the same scale. We take $m_1 = m_2 = m$ for simplicity (and thus $\vAcm=v_{2{\rm,cm}}=v$), and we fix the elastic scattering cross section $ \sel = \bar \sigma = (100\mev)^{-2}$ by the loose argument above. Observations then demand $m \sim 200 \mev$, but observables in the velocity range of interest are set by $\bar \sigma b^k/m$ (where $k=1/2,1$ in the similar-mass or the massless fusion partner scenario, respectively), which we comment on in depth below. In the left panel of \Fig{plot} we assume that all reactants have similar mass, and the velocity dependence of the fusion cross section goes like $\sff \vAcm \simeq (2b+\vAcm^2)^{1/2} \bar \sigma.$ At high velocity, the fusion and elastic cross sections coincide. In the right panel of \Fig{plot} we assume that one outgoing fusion product is massless, so that the velocity dependence of fusion goes like $\sff \vAcm \simeq (2b+\vAcm^2) \bar \sigma.$ Fusion becomes negligible at high velocity, so elastic scattering determines the structure of larger astrophysical objects. In both panels, we display a limit $\bar \sigma \leq 0.4\cm^2 \!/\g$ from stacked observations of cluster mergers at $v= 1000 \pm 200\km\!/\s$ \cite{Harvey:2015hha}, which is now thought to be stronger than the limit from mass loss in the Bullet Cluster \cite{Kim:2016ujt}. Because observations favor $(\sigma v/m)|_{\rm cluster} \simeq (\sigma v/m)|_{\rm dwarf}$, we see that fusion must dominate at and below the cluster scale. This implies $b \gtrsim v_{\rm cluster}^{1/k}/2$: in other words, $b \gtrsim 10^{-6}, 10^{-3}$ depending on the dark sector nuclear physics.

At low velocity, the same results will obtain for any rescaling of $\bar \sigma$, $b$, and $m$ that keeps $\bar \sigma b^k/m$ fixed. For $m \gtrsim \cO(10\mev)$ the dark matter particles are out of thermal equilibrium at the time of BBN, so we need not specify a dark sector thermal history to avoid running afoul of degree of freedom counting at that epoch \cite{Cyburt:2015mya}. Compatibility with Standard Model BBN could provide bounds on the dark sector if energy injection at that time is large \cite{Fradette:2017sdd}, but this will not pose a problem if the final state momenta are small (which is determined as in \Eqs{momenta-NR}{momenta-A40} by $b$ and $m$) or if the dark sector is absolutely decoupled from the SM (which can be true to arbitrary accuracy, since we only need {\it self}-interactions to thermalize the cores of halos). If we take the limit $b\to 1$, ``fusion'' becomes annihilation \cite{Kaplinghat:2000vt}. The small-$b$ limit worked out explicitly above is changed at the $\cO(1)$ level, but a similar analysis follows. Since $b$ is bounded from above, we find that we may still fit observations with $m$ as large as $\sim \few \gev$ for fixed $\bar \sigma/m$ using $\bar \sigma v = 4\pi/m^2$, or larger but with a different velocity dependence if $\bar \sigma$ saturates perturbative limits \cite{Hui:2001wy}. However, the scale of \Eq{SIDM-favored} is na\"ively too large for annihilation or semi-annihilation, since antiparticles are necessary fuel and they are exponentially depleted during freezeout if the annihilation cross section is as large as \Eq{SIDM-favored} indicates. Model building solutions exist \cite{Kaplinghat:2000vt} but are constrained by non-observation of annihilation products \cite{Beacom:2006tt}. Dark fusion provides a ``gentler'' alternative.

The fusion reaction releases pent-up energy $b(m_1+m_2)$, so the momentum imparted to the fusion products could allow them to evaporate from shallow gravitational potential wells. This would serve as an attractive alternative realization of the scenario of \cite{Wang:2014ina}, which sought to give dark matter particles a nonrelativistic kick via a compressed decay. Such kicks allow structures below certain mass scales to be tidally disrupted, potentially alleviating concerns about the number of collapsed substructures in the Milky Way vicinity \cite{Klypin:1999uc, BoylanKolchin:2011de} (but also see \cite{Jiang:2015vra}). These kicks inevitably turn on in the late universe when densities are sufficiently large that the fusion rate $n \sff \vAcm$ is comparable to a gravitational infall time $1/\sqrt{G \rho}$, rather than turning on at a lifetime which is coincidentally comparable to the current Hubble scale.

We emphasize that this model predicts measurements of $\sigma v$ with a very different dependence on host mass compared to the model of \cite{Kaplinghat:2015aga, Tulin:2012wi, Tulin:2013teo}. Dark fusion has an exactly flat $\sigma v$ until $\vAcm \gtrsim (2bm_2/m_1)^k$, where $k=1/2$ for similar-mass final states and $k=1$ for a massless fusion partner. If the final state particles have similar masses, the fusion and elastic scattering cross sections become similar above this threshold. If one final state particle is very light, fusion becomes negligible above this threshold, and elastic scatterings dominate. If dark matter chemistry varies between local environments, it is possible that there will be a family of curves populating the $\langle \sigma v \rangle-v$ plot at high velocities. However, a constant $\sff v$ down to arbitrarily small host mass is a robust prediction of this model. This is qualitatively different than the dark photon model of \cite{Kaplinghat:2015aga, Tulin:2012wi, Tulin:2013teo}, which relies on elastic scattering only. Models with a dark photon of mass $m_\phi$ have a cross section that varies strongly with velocity: one finds $\sigma_s v \sim v$ in small hosts with low velocity dispersions, because interactions are contact-like; then a maximum in objects with characteristic momentum transfer $m v \sim m_\phi$, where a cancellation occurs in the denominator of $\sigma_s$; and a scaling like $\sigma_s v \sim v^{-4}$ in large halos, when $m_\phi$ is much less than the momentum transfer. Dark fusion is also qualitatively different from an inelastic process with a threshold, like exciting dark matter \cite{Finkbeiner:2007kk}, which has a high-velocity threshold below which inelasticity turns off.

Detailed observations of the inner structure of a virialized object with $v \simeq 700 \pm 100 \km/\s$ or $v \ll 30\km/\s$ could differentiate the dark fusion and dark photon models. One interesting candidate for this purpose is PGC 43296, identified in 2MASS Redshift Survey data \cite{Huchra:2011aa}. This galaxy group has 191 members, mass $4.2\tenx{14}M_\odot \leq5.2\tenx{14}M_\odot$, and a velocity dispersion $595 \km/\s \leq \sigma \leq 800 \km/\s$, with 3\% systematic error bars \cite{Kourkchi:2017aa}. Near this velocity, the dark photon model has a maximum $\sff \vAcm$ while dark fusion predicts the same $\sff \vAcm$ as in dwarf galaxies. The dark photon model would lead to a significantly larger core and {\it lower} stellar velocities inside of the radius where dark matter scattering takes place once per Hubble time.

Finally, we point out that dark fusion may be accompanied by indirect detection signals if the dark sector is coupled even weakly to the Standard Model \cite{Detmold:2014qqa}. Semi-annihilation is the $b\sim 0.5$ limit of dark fusion with a massless partner, but with cross section as in \Eq{SIDM-favored} would lead to higher rates and lower mass scales than previously considered \cite{Boehm:2014bia, Ko:2014gha}. This could lead to interesting signals or strong limits on the decays of the outgoing fusion partner.

\section{Early Universe Cosmology of Models with Dark Fusion}
\label{sec:models}
Because the required cross section \Eq{SIDM-favored} is so large, we do not discuss symmetric thermal relics \cite{Kaplinghat:2000vt}. Two-to-two fusion can be thought of as very compressed annihilation without antiparticles, although it does require some fuel to survive from the early universe if it operates at late times. We comment here on particle physics models that could provide the ingredients necessary for dark fusion to be operative in small-scale structures in the present epoch. The model requirements are that two-to-two reactions are dominant, unbound nucleons are abundant, and reactions are nonrelativistic so that the $\sff \vAcm \propto \vAcm^0$ scaling holds \footnote{We thank Moira Gresham, Tim Lou, and Kathryn Zurek for emphasizing these essential points.}. As noted above, nucleosynthesis happens at temperatures with $v^2\sim\cO(1/20-1/200)$ \cite{Detmold:2014qqa} while galactic dark fusion happens with velocities $v \lesssim 10^{-2}$, so it is plausible that different chemical reactions will dominate in the two epochs and that unspent fuel remains in galaxies today. No explicit computation yet manifests all the features of interest, but many compelling models are just beginning to be explored. A complete model that results in a nontrivial early universe yield of fusion participants with a cross section of the size of \Eq{SIDM-favored} is hopefully within model-building reach.

In the model of \cite{Krnjaic:2014xza}, ``dark protons'' are the only matter. They have a small charge under a dark Coulomb force as well as a stronger nuclear force that is the residual of a dark SU($N$). The binding energy based on the liquid drop model is \cite{Baym:1971ax, Mackie:1977nof}
\begin{align} \label{liq-drop}
\cB(A) &= a_V A - a_S A^{2/3} - a_C A^{2-1/3} \pm \frac{a_P}{ A^{1/2}}.
\end{align}
The Coulomb coefficient $a_C$ should be smaller than the strong-binding coefficients. Taking \Eq{liq-drop} literally, partial fission of states $2A_i \to A_{i-1}+A_{i+1}$, as in our explicit example in \Fig{plot}, is weakly exothermic for $i \gtrsim 20$ when $ \alpha_C/\alpha_s = a_C/a_V = a_C/a_S = 1/100$, where $\alpha_{C,s}$ are the dark Coulomb and strong fine structure constants. The reverse reaction $ A_{i-1}+A_{i+1} \to  2A_i $ is (more strongly) exothermic at smaller $i$. The early universe yields of dark nuclei will determine which kinds of processes are dominant in galaxies today. If ``strongly exothermic'' processes determine the early universe yield and most particles wind up with $A \sim 20$, as hinted in \cite{Krnjaic:2014xza}, then it is reasonable to expect that dark fusion today will be in the weakly exothermic regime. The liquid drop model is best interpreted as an ans\"atz, however, and more complete calculations of the nuclear binding are desirable \cite{Gresham:2017zqi, Gresham:2017cvl}.

In \cite{Detmold:2014qqa}, the spectrum of a strongly coupled two-color, two-flavor QCD-like model is presented. The authors find a light scalar $\pi_d$ and a light vector $\rho_d$ that can fuse into a nucleus $D$ by emitting a light dark Higgs $h_d$. The particle masses are near the dark QCD scale with the exception of $h_d$, which is lighter; heavier three- and four-body states may be stable as well \cite{Detmold:2014kba}. A nonnegligible yield of all species is obtained for a wide range of dark sector cross sections, which serves as a nontrivial demonstration of the potential relevance of our mechanism for late times. Because the yield of dark sector particles leads to significant quantities of each species \cite{Detmold:2014qqa} but cross sections as large as \Eq{SIDM-favored} have not yet been explored, this model is highly motivated to more explicitly investigate in the context of small-scale structure.

Models with no dark-sector fermions and only glueballs above a mass gap \cite{Forestell:2016qhc, Forestell:2017wov} are also candidates for late-time dark fusion. Inelastic two-to-two scattering of excited glueball states to lower-lying ones is a candidate mechanism with similar features to dark fusion, but the amount of binding energy released would need to be fairly large compared to na\"ive expectation.

Models with matter that is coupled to an attractive Yukawa force only \cite{Wise:2014jva, Wise:2014ola, Gresham:2017zqi, Gresham:2017cvl} are also intriguing. These models lead to composite states of extremely large dark nucleon number, aggregating enormous quantities of particles in the early universe. Whether the late-time cosmology of these ``nuggets'' can be dominated by the simple cross sections worked out here is unclear and deserves further study. In particular, it is unclear if local fusion rates, fractional binding energy release, and particle multiplicities can match on to the simple velocity dependence suggested here \cite{Note2}.

Finally, models with a dissipative dark component \cite{Fan:2013yva, Fan:2013tia, McCullough:2013jma, Boddy:2014yra, Boddy:2016bbu, Agrawal:2016quu} have the necessary ingredients to undergo dark fusion. The late-universe cosmology of such models is an area of active research, and we simply comment here that inelasticities, including fusion as well as hyperfine transitions and decays, could be qualitatively linked in determining the late-time evolution of the inner structure of these halos.

\section{Conclusions}
\label{sec:conc}

In this paper, we have proposed that the cross section times velocity for dark matter self-interactions may have no apparent velocity dependence in the late universe because the dark matter is undergoing two-to-two fusion. In \Fig{plot}, we show that this scenario would explain the evident flatness of dark matter cross section times velocity over many decades in velocity that was proposed as an explanation for cores in dark matter halos of widely varying masses \cite{Kaplinghat:2015aga, Kamada:2016euw}. It may also provide nonrelativistic kicks to fusion daughters that can evaporate small, loosely bound dark matter substructures \cite{Wang:2014ina}.

This model provides a firm prediction: {\it observed galaxy cores will point to the same value of $\sigma v$ for all velocities below cluster scales}. There will be a linear rise in the required value of $\sigma v$ at higher velocities, either due to the high-velocity regime in \Eq{flux-func} (for the similar-mass scenario) or due to dominance of elastic scatterings at high velocities (in the massless-fusion-partner case). This differs markedly from $\sigma v$ at the same scales in the dark photon model of \cite{Kaplinghat:2015aga}. Thus, improved astrophysical observations at low, intermediate, or high masses could shed light on the details of the dark sector. We have noted the relevance of galaxy group PGC 43296 \cite{Kourkchi:2017aa} for breaking the degeneracy between these models. It is also possible that astrophysical objects have different chemical compositions, with the linear rise of $\sigma v$ turning on at different values of the velocity in different hosts.  We may discover hints of dark fusion activity as we look deeper into dark matter potential wells!

~\\~ \noindent {\bf Acknowledgments:} Fermilab is operated by Fermi Research Alliance, LLC under Contract No. De-AC02-07CH11359 with the United States Department of Energy. SDM gratefully acknowledges discussion with Nick Gnedin, Dan Hooper, Gordan Krnjaic, and Haibo Yu (who provided the points and the curve from Fig.~1 of \cite{Kaplinghat:2015aga}); encouragement from Zackaria Chacko, Tim Cohen, and Rouven Essig; motivation from Sean Tulin (who posed the basic question at the Aspen Center for Physics, which is supported by National Science Foundation grant PHY-1066293); and essential help from Susan Gardner (who he deeply thanks for pointing to and clearly explaining \cite{Bethe:1950jm}, and without whom this paper would not have been written) as well as Moira Gresham, Tim Lou, and Kathryn Zurek (who explained some key features of \cite{Gresham:2017zqi, Gresham:2017cvl}). SDM thanks various hosts at Kentucky, BU, Berkeley, SLAC, UO, and UW for hospitality.

\bibliography{late-dark-fusion}

\begin{thebibliography}{44}%
\makeatletter
\providecommand \@ifxundefined [1]{%
 \@ifx{#1\undefined}
}%
\providecommand \@ifnum [1]{%
 \ifnum #1\expandafter \@firstoftwo
 \else \expandafter \@secondoftwo
 \fi
}%
\providecommand \@ifx [1]{%
 \ifx #1\expandafter \@firstoftwo
 \else \expandafter \@secondoftwo
 \fi
}%
\providecommand \natexlab [1]{#1}%
\providecommand \enquote  [1]{``#1''}%
\providecommand \bibnamefont  [1]{#1}%
\providecommand \bibfnamefont [1]{#1}%
\providecommand \citenamefont [1]{#1}%
\providecommand \href@noop [0]{\@secondoftwo}%
\providecommand \href [0]{\begingroup \@sanitize@url \@href}%
\providecommand \@href[1]{\@@startlink{#1}\@@href}%
\providecommand \@@href[1]{\endgroup#1\@@endlink}%
\providecommand \@sanitize@url [0]{\catcode `\\12\catcode `\$12\catcode
  `\&12\catcode `\#12\catcode `\^12\catcode `\_12\catcode `\%12\relax}%
\providecommand \@@startlink[1]{}%
\providecommand \@@endlink[0]{}%
\providecommand \url  [0]{\begingroup\@sanitize@url \@url }%
\providecommand \@url [1]{\endgroup\@href {#1}{\urlprefix }}%
\providecommand \urlprefix  [0]{URL }%
\providecommand \Eprint [0]{\href }%
\providecommand \doibase [0]{http://dx.doi.org/}%
\providecommand \selectlanguage [0]{\@gobble}%
\providecommand \bibinfo  [0]{\@secondoftwo}%
\providecommand \bibfield  [0]{\@secondoftwo}%
\providecommand \translation [1]{[#1]}%
\providecommand \BibitemOpen [0]{}%
\providecommand \bibitemStop [0]{}%
\providecommand \bibitemNoStop [0]{.\EOS\space}%
\providecommand \EOS [0]{\spacefactor3000\relax}%
\providecommand \BibitemShut  [1]{\csname bibitem#1\endcsname}%
\let\auto@bib@innerbib\@empty
\bibitem [{\citenamefont {Bullock}\ and\ \citenamefont
  {Boylan-Kolchin}(2017)}]{Bullock:2017xww}%
  \BibitemOpen
  \bibfield  {author} {\bibinfo {author} {\bibfnamefont {J.~S.}\ \bibnamefont
  {Bullock}}\ and\ \bibinfo {author} {\bibfnamefont {M.}~\bibnamefont
  {Boylan-Kolchin}},\ }\href {\doibase 10.1146/annurev-astro-091916-055313}
  {\bibfield  {journal} {\bibinfo  {journal} {Ann. Rev. Astron. Astrophys.}\
  }\textbf {\bibinfo {volume} {55}},\ \bibinfo {pages} {343} (\bibinfo {year}
  {2017})},\ \Eprint {http://arxiv.org/abs/1707.04256} {arXiv:1707.04256
  [astro-ph.CO]} \BibitemShut {NoStop}%
\bibitem [{\citenamefont {Tulin}\ and\ \citenamefont
  {Yu}(2017)}]{Tulin:2017ara}%
  \BibitemOpen
  \bibfield  {author} {\bibinfo {author} {\bibfnamefont {S.}~\bibnamefont
  {Tulin}}\ and\ \bibinfo {author} {\bibfnamefont {H.-B.}\ \bibnamefont {Yu}},\
  }\href@noop {} {\  (\bibinfo {year} {2017})},\ \Eprint
  {http://arxiv.org/abs/1705.02358} {arXiv:1705.02358 [hep-ph]} \BibitemShut
  {NoStop}%
\bibitem [{\citenamefont {Kaplinghat}\ \emph {et~al.}(2016)\citenamefont
  {Kaplinghat}, \citenamefont {Tulin},\ and\ \citenamefont
  {Yu}}]{Kaplinghat:2015aga}%
  \BibitemOpen
  \bibfield  {author} {\bibinfo {author} {\bibfnamefont {M.}~\bibnamefont
  {Kaplinghat}}, \bibinfo {author} {\bibfnamefont {S.}~\bibnamefont {Tulin}}, \
  and\ \bibinfo {author} {\bibfnamefont {H.-B.}\ \bibnamefont {Yu}},\ }\href
  {\doibase 10.1103/PhysRevLett.116.041302} {\bibfield  {journal} {\bibinfo
  {journal} {Phys. Rev. Lett.}\ }\textbf {\bibinfo {volume} {116}},\ \bibinfo
  {pages} {041302} (\bibinfo {year} {2016})},\ \Eprint
  {http://arxiv.org/abs/1508.03339} {arXiv:1508.03339 [astro-ph.CO]}
  \BibitemShut {NoStop}%
\bibitem [{\citenamefont {Kamada}\ \emph {et~al.}(2017)\citenamefont {Kamada},
  \citenamefont {Kaplinghat}, \citenamefont {Pace},\ and\ \citenamefont
  {Yu}}]{Kamada:2016euw}%
  \BibitemOpen
  \bibfield  {author} {\bibinfo {author} {\bibfnamefont {A.}~\bibnamefont
  {Kamada}}, \bibinfo {author} {\bibfnamefont {M.}~\bibnamefont {Kaplinghat}},
  \bibinfo {author} {\bibfnamefont {A.~B.}\ \bibnamefont {Pace}}, \ and\
  \bibinfo {author} {\bibfnamefont {H.-B.}\ \bibnamefont {Yu}},\ }\href
  {\doibase 10.1103/PhysRevLett.119.111102} {\bibfield  {journal} {\bibinfo
  {journal} {Phys. Rev. Lett.}\ }\textbf {\bibinfo {volume} {119}},\ \bibinfo
  {pages} {111102} (\bibinfo {year} {2017})},\ \Eprint
  {http://arxiv.org/abs/1611.02716} {arXiv:1611.02716 [astro-ph.GA]}
  \BibitemShut {NoStop}%
\bibitem [{\citenamefont {Krnjaic}\ and\ \citenamefont
  {Sigurdson}(2015)}]{Krnjaic:2014xza}%
  \BibitemOpen
  \bibfield  {author} {\bibinfo {author} {\bibfnamefont {G.}~\bibnamefont
  {Krnjaic}}\ and\ \bibinfo {author} {\bibfnamefont {K.}~\bibnamefont
  {Sigurdson}},\ }\href {\doibase 10.1016/j.physletb.2015.11.001} {\bibfield
  {journal} {\bibinfo  {journal} {Phys. Lett.}\ }\textbf {\bibinfo {volume}
  {B751}},\ \bibinfo {pages} {464} (\bibinfo {year} {2015})},\ \Eprint
  {http://arxiv.org/abs/1406.1171} {arXiv:1406.1171 [hep-ph]} \BibitemShut
  {NoStop}%
\bibitem [{\citenamefont {Detmold}\ \emph
  {et~al.}(2014{\natexlab{a}})\citenamefont {Detmold}, \citenamefont
  {McCullough},\ and\ \citenamefont {Pochinsky}}]{Detmold:2014qqa}%
  \BibitemOpen
  \bibfield  {author} {\bibinfo {author} {\bibfnamefont {W.}~\bibnamefont
  {Detmold}}, \bibinfo {author} {\bibfnamefont {M.}~\bibnamefont {McCullough}},
  \ and\ \bibinfo {author} {\bibfnamefont {A.}~\bibnamefont {Pochinsky}},\
  }\href {\doibase 10.1103/PhysRevD.90.115013} {\bibfield  {journal} {\bibinfo
  {journal} {Phys. Rev.}\ }\textbf {\bibinfo {volume} {D90}},\ \bibinfo {pages}
  {115013} (\bibinfo {year} {2014}{\natexlab{a}})},\ \Eprint
  {http://arxiv.org/abs/1406.2276} {arXiv:1406.2276 [hep-ph]} \BibitemShut
  {NoStop}%
\bibitem [{\citenamefont {Wise}\ and\ \citenamefont
  {Zhang}(2014)}]{Wise:2014jva}%
  \BibitemOpen
  \bibfield  {author} {\bibinfo {author} {\bibfnamefont {M.~B.}\ \bibnamefont
  {Wise}}\ and\ \bibinfo {author} {\bibfnamefont {Y.}~\bibnamefont {Zhang}},\
  }\href {\doibase 10.1103/PhysRevD.90.055030, 10.1103/PhysRevD.91.039907}
  {\bibfield  {journal} {\bibinfo  {journal} {Phys. Rev.}\ }\textbf {\bibinfo
  {volume} {D90}},\ \bibinfo {pages} {055030} (\bibinfo {year} {2014})},\
  \bibinfo {note} {[Erratum: Phys. Rev.D91,no.3,039907(2015)]},\ \Eprint
  {http://arxiv.org/abs/1407.4121} {arXiv:1407.4121 [hep-ph]} \BibitemShut
  {NoStop}%
\bibitem [{\citenamefont {Wise}\ and\ \citenamefont
  {Zhang}(2015)}]{Wise:2014ola}%
  \BibitemOpen
  \bibfield  {author} {\bibinfo {author} {\bibfnamefont {M.~B.}\ \bibnamefont
  {Wise}}\ and\ \bibinfo {author} {\bibfnamefont {Y.}~\bibnamefont {Zhang}},\
  }\href {\doibase 10.1007/JHEP10(2015)165, 10.1007/JHEP02(2015)023} {\bibfield
   {journal} {\bibinfo  {journal} {JHEP}\ }\textbf {\bibinfo {volume} {02}},\
  \bibinfo {pages} {023} (\bibinfo {year} {2015})},\ \bibinfo {note} {[Erratum:
  JHEP10,165(2015)]},\ \Eprint {http://arxiv.org/abs/1411.1772}
  {arXiv:1411.1772 [hep-ph]} \BibitemShut {NoStop}%
\bibitem [{\citenamefont {Hardy}\ \emph {et~al.}(2015)\citenamefont {Hardy},
  \citenamefont {Lasenby}, \citenamefont {March-Russell},\ and\ \citenamefont
  {West}}]{Hardy:2014mqa}%
  \BibitemOpen
  \bibfield  {author} {\bibinfo {author} {\bibfnamefont {E.}~\bibnamefont
  {Hardy}}, \bibinfo {author} {\bibfnamefont {R.}~\bibnamefont {Lasenby}},
  \bibinfo {author} {\bibfnamefont {J.}~\bibnamefont {March-Russell}}, \ and\
  \bibinfo {author} {\bibfnamefont {S.~M.}\ \bibnamefont {West}},\ }\href
  {\doibase 10.1007/JHEP06(2015)011} {\bibfield  {journal} {\bibinfo  {journal}
  {JHEP}\ }\textbf {\bibinfo {volume} {06}},\ \bibinfo {pages} {011} (\bibinfo
  {year} {2015})},\ \Eprint {http://arxiv.org/abs/1411.3739} {arXiv:1411.3739
  [hep-ph]} \BibitemShut {NoStop}%
\bibitem [{\citenamefont {Forestell}\ \emph
  {et~al.}(2017{\natexlab{a}})\citenamefont {Forestell}, \citenamefont
  {Morrissey},\ and\ \citenamefont {Sigurdson}}]{Forestell:2016qhc}%
  \BibitemOpen
  \bibfield  {author} {\bibinfo {author} {\bibfnamefont {L.}~\bibnamefont
  {Forestell}}, \bibinfo {author} {\bibfnamefont {D.~E.}\ \bibnamefont
  {Morrissey}}, \ and\ \bibinfo {author} {\bibfnamefont {K.}~\bibnamefont
  {Sigurdson}},\ }\href {\doibase 10.1103/PhysRevD.95.015032} {\bibfield
  {journal} {\bibinfo  {journal} {Phys. Rev.}\ }\textbf {\bibinfo {volume}
  {D95}},\ \bibinfo {pages} {015032} (\bibinfo {year} {2017}{\natexlab{a}})},\
  \Eprint {http://arxiv.org/abs/1605.08048} {arXiv:1605.08048 [hep-ph]}
  \BibitemShut {NoStop}%
\bibitem [{\citenamefont {Gresham}\ \emph
  {et~al.}(2017{\natexlab{a}})\citenamefont {Gresham}, \citenamefont {Lou},\
  and\ \citenamefont {Zurek}}]{Gresham:2017cvl}%
  \BibitemOpen
  \bibfield  {author} {\bibinfo {author} {\bibfnamefont {M.~I.}\ \bibnamefont
  {Gresham}}, \bibinfo {author} {\bibfnamefont {H.~K.}\ \bibnamefont {Lou}}, \
  and\ \bibinfo {author} {\bibfnamefont {K.~M.}\ \bibnamefont {Zurek}},\
  }\href@noop {} {\  (\bibinfo {year} {2017}{\natexlab{a}})},\ \Eprint
  {http://arxiv.org/abs/1707.02316} {arXiv:1707.02316 [hep-ph]} \BibitemShut
  {NoStop}%
\bibitem [{\citenamefont {Gresham}\ \emph
  {et~al.}(2017{\natexlab{b}})\citenamefont {Gresham}, \citenamefont {Lou},\
  and\ \citenamefont {Zurek}}]{Gresham:2017zqi}%
  \BibitemOpen
  \bibfield  {author} {\bibinfo {author} {\bibfnamefont {M.~I.}\ \bibnamefont
  {Gresham}}, \bibinfo {author} {\bibfnamefont {H.~K.}\ \bibnamefont {Lou}}, \
  and\ \bibinfo {author} {\bibfnamefont {K.~M.}\ \bibnamefont {Zurek}},\
  }\href@noop {} {\  (\bibinfo {year} {2017}{\natexlab{b}})},\ \Eprint
  {http://arxiv.org/abs/1707.02313} {arXiv:1707.02313 [hep-ph]} \BibitemShut
  {NoStop}%
\bibitem [{\citenamefont {Forestell}\ \emph
  {et~al.}(2017{\natexlab{b}})\citenamefont {Forestell}, \citenamefont
  {Morrissey},\ and\ \citenamefont {Sigurdson}}]{Forestell:2017wov}%
  \BibitemOpen
  \bibfield  {author} {\bibinfo {author} {\bibfnamefont {L.}~\bibnamefont
  {Forestell}}, \bibinfo {author} {\bibfnamefont {D.~E.}\ \bibnamefont
  {Morrissey}}, \ and\ \bibinfo {author} {\bibfnamefont {K.}~\bibnamefont
  {Sigurdson}},\ }\href@noop {} {\  (\bibinfo {year} {2017}{\natexlab{b}})},\
  \Eprint {http://arxiv.org/abs/1710.06447} {arXiv:1710.06447 [hep-ph]}
  \BibitemShut {NoStop}%
\bibitem [{\citenamefont {Tulin}\ \emph
  {et~al.}(2013{\natexlab{a}})\citenamefont {Tulin}, \citenamefont {Yu},\ and\
  \citenamefont {Zurek}}]{Tulin:2012wi}%
  \BibitemOpen
  \bibfield  {author} {\bibinfo {author} {\bibfnamefont {S.}~\bibnamefont
  {Tulin}}, \bibinfo {author} {\bibfnamefont {H.-B.}\ \bibnamefont {Yu}}, \
  and\ \bibinfo {author} {\bibfnamefont {K.~M.}\ \bibnamefont {Zurek}},\ }\href
  {\doibase 10.1103/PhysRevLett.110.111301} {\bibfield  {journal} {\bibinfo
  {journal} {Phys. Rev. Lett.}\ }\textbf {\bibinfo {volume} {110}},\ \bibinfo
  {pages} {111301} (\bibinfo {year} {2013}{\natexlab{a}})},\ \Eprint
  {http://arxiv.org/abs/1210.0900} {arXiv:1210.0900 [hep-ph]} \BibitemShut
  {NoStop}%
\bibitem [{\citenamefont {Tulin}\ \emph
  {et~al.}(2013{\natexlab{b}})\citenamefont {Tulin}, \citenamefont {Yu},\ and\
  \citenamefont {Zurek}}]{Tulin:2013teo}%
  \BibitemOpen
  \bibfield  {author} {\bibinfo {author} {\bibfnamefont {S.}~\bibnamefont
  {Tulin}}, \bibinfo {author} {\bibfnamefont {H.-B.}\ \bibnamefont {Yu}}, \
  and\ \bibinfo {author} {\bibfnamefont {K.~M.}\ \bibnamefont {Zurek}},\ }\href
  {\doibase 10.1103/PhysRevD.87.115007} {\bibfield  {journal} {\bibinfo
  {journal} {Phys. Rev.}\ }\textbf {\bibinfo {volume} {D87}},\ \bibinfo {pages}
  {115007} (\bibinfo {year} {2013}{\natexlab{b}})},\ \Eprint
  {http://arxiv.org/abs/1302.3898} {arXiv:1302.3898 [hep-ph]} \BibitemShut
  {NoStop}%
\bibitem [{\citenamefont {Hui}(2001)}]{Hui:2001wy}%
  \BibitemOpen
  \bibfield  {author} {\bibinfo {author} {\bibfnamefont {L.}~\bibnamefont
  {Hui}},\ }\href {\doibase 10.1103/PhysRevLett.86.3467} {\bibfield  {journal}
  {\bibinfo  {journal} {Phys. Rev. Lett.}\ }\textbf {\bibinfo {volume} {86}},\
  \bibinfo {pages} {3467} (\bibinfo {year} {2001})},\ \Eprint
  {http://arxiv.org/abs/astro-ph/0102349} {arXiv:astro-ph/0102349 [astro-ph]}
  \BibitemShut {NoStop}%
\bibitem [{\citenamefont {Kaplinghat}\ \emph {et~al.}(2000)\citenamefont
  {Kaplinghat}, \citenamefont {Knox},\ and\ \citenamefont
  {Turner}}]{Kaplinghat:2000vt}%
  \BibitemOpen
  \bibfield  {author} {\bibinfo {author} {\bibfnamefont {M.}~\bibnamefont
  {Kaplinghat}}, \bibinfo {author} {\bibfnamefont {L.}~\bibnamefont {Knox}}, \
  and\ \bibinfo {author} {\bibfnamefont {M.~S.}\ \bibnamefont {Turner}},\
  }\href {\doibase 10.1103/PhysRevLett.85.3335} {\bibfield  {journal} {\bibinfo
   {journal} {Phys. Rev. Lett.}\ }\textbf {\bibinfo {volume} {85}},\ \bibinfo
  {pages} {3335} (\bibinfo {year} {2000})},\ \Eprint
  {http://arxiv.org/abs/astro-ph/0005210} {arXiv:astro-ph/0005210 [astro-ph]}
  \BibitemShut {NoStop}%
\bibitem [{\citenamefont {Beacom}\ \emph {et~al.}(2007)\citenamefont {Beacom},
  \citenamefont {Bell},\ and\ \citenamefont {Mack}}]{Beacom:2006tt}%
  \BibitemOpen
  \bibfield  {author} {\bibinfo {author} {\bibfnamefont {J.~F.}\ \bibnamefont
  {Beacom}}, \bibinfo {author} {\bibfnamefont {N.~F.}\ \bibnamefont {Bell}}, \
  and\ \bibinfo {author} {\bibfnamefont {G.~D.}\ \bibnamefont {Mack}},\ }\href
  {\doibase 10.1103/PhysRevLett.99.231301} {\bibfield  {journal} {\bibinfo
  {journal} {Phys. Rev. Lett.}\ }\textbf {\bibinfo {volume} {99}},\ \bibinfo
  {pages} {231301} (\bibinfo {year} {2007})},\ \Eprint
  {http://arxiv.org/abs/astro-ph/0608090} {arXiv:astro-ph/0608090 [astro-ph]}
  \BibitemShut {NoStop}%
\bibitem [{\citenamefont {Boehm}\ \emph {et~al.}(2014)\citenamefont {Boehm},
  \citenamefont {Dolan},\ and\ \citenamefont {McCabe}}]{Boehm:2014bia}%
  \BibitemOpen
  \bibfield  {author} {\bibinfo {author} {\bibfnamefont {C.}~\bibnamefont
  {Boehm}}, \bibinfo {author} {\bibfnamefont {M.~J.}\ \bibnamefont {Dolan}}, \
  and\ \bibinfo {author} {\bibfnamefont {C.}~\bibnamefont {McCabe}},\ }\href
  {\doibase 10.1103/PhysRevD.90.023531} {\bibfield  {journal} {\bibinfo
  {journal} {Phys. Rev.}\ }\textbf {\bibinfo {volume} {D90}},\ \bibinfo {pages}
  {023531} (\bibinfo {year} {2014})},\ \Eprint {http://arxiv.org/abs/1404.4977}
  {arXiv:1404.4977 [hep-ph]} \BibitemShut {NoStop}%
\bibitem [{\citenamefont {Ko}\ \emph {et~al.}(2014)\citenamefont {Ko},
  \citenamefont {Park},\ and\ \citenamefont {Tang}}]{Ko:2014gha}%
  \BibitemOpen
  \bibfield  {author} {\bibinfo {author} {\bibfnamefont {P.}~\bibnamefont
  {Ko}}, \bibinfo {author} {\bibfnamefont {W.-I.}\ \bibnamefont {Park}}, \ and\
  \bibinfo {author} {\bibfnamefont {Y.}~\bibnamefont {Tang}},\ }\href {\doibase
  10.1088/1475-7516/2014/09/013} {\bibfield  {journal} {\bibinfo  {journal}
  {JCAP}\ }\textbf {\bibinfo {volume} {1409}},\ \bibinfo {pages} {013}
  (\bibinfo {year} {2014})},\ \Eprint {http://arxiv.org/abs/1404.5257}
  {arXiv:1404.5257 [hep-ph]} \BibitemShut {NoStop}%
\bibitem [{\citenamefont {Harvey}\ \emph {et~al.}(2015)\citenamefont {Harvey},
  \citenamefont {Massey}, \citenamefont {Kitching}, \citenamefont {Taylor},\
  and\ \citenamefont {Tittley}}]{Harvey:2015hha}%
  \BibitemOpen
  \bibfield  {author} {\bibinfo {author} {\bibfnamefont {D.}~\bibnamefont
  {Harvey}}, \bibinfo {author} {\bibfnamefont {R.}~\bibnamefont {Massey}},
  \bibinfo {author} {\bibfnamefont {T.}~\bibnamefont {Kitching}}, \bibinfo
  {author} {\bibfnamefont {A.}~\bibnamefont {Taylor}}, \ and\ \bibinfo {author}
  {\bibfnamefont {E.}~\bibnamefont {Tittley}},\ }\href {\doibase
  10.1126/science.1261381} {\bibfield  {journal} {\bibinfo  {journal}
  {Science}\ }\textbf {\bibinfo {volume} {347}},\ \bibinfo {pages} {1462}
  (\bibinfo {year} {2015})},\ \Eprint {http://arxiv.org/abs/1503.07675}
  {arXiv:1503.07675 [astro-ph.CO]} \BibitemShut {NoStop}%
\bibitem [{\citenamefont {Bethe}\ and\ \citenamefont
  {Longmire}(1950)}]{Bethe:1950jm}%
  \BibitemOpen
  \bibfield  {author} {\bibinfo {author} {\bibfnamefont {H.~A.}\ \bibnamefont
  {Bethe}}\ and\ \bibinfo {author} {\bibfnamefont {C.}~\bibnamefont
  {Longmire}},\ }\href {\doibase 10.1103/PhysRev.77.647} {\bibfield  {journal}
  {\bibinfo  {journal} {Phys. Rev.}\ }\textbf {\bibinfo {volume} {77}},\
  \bibinfo {pages} {647} (\bibinfo {year} {1950})}\BibitemShut {NoStop}%
\bibitem [{Note1()}]{Note1}%
  \BibitemOpen
  \bibinfo {note} {We thank Susan Gardner for emphasizing the importance and
  generality of this result.}\BibitemShut {Stop}%
\bibitem [{\citenamefont {Kim}\ \emph {et~al.}(2017)\citenamefont {Kim},
  \citenamefont {Peter},\ and\ \citenamefont {Wittman}}]{Kim:2016ujt}%
  \BibitemOpen
  \bibfield  {author} {\bibinfo {author} {\bibfnamefont {S.~Y.}\ \bibnamefont
  {Kim}}, \bibinfo {author} {\bibfnamefont {A.~H.~G.}\ \bibnamefont {Peter}}, \
  and\ \bibinfo {author} {\bibfnamefont {D.}~\bibnamefont {Wittman}},\ }\href
  {\doibase 10.1093/mnras/stx896} {\bibfield  {journal} {\bibinfo  {journal}
  {Mon. Not. Roy. Astron. Soc.}\ }\textbf {\bibinfo {volume} {469}},\ \bibinfo
  {pages} {1414} (\bibinfo {year} {2017})},\ \Eprint
  {http://arxiv.org/abs/1608.08630} {arXiv:1608.08630 [astro-ph.CO]}
  \BibitemShut {NoStop}%
\bibitem [{\citenamefont {Cyburt}\ \emph {et~al.}(2016)\citenamefont {Cyburt},
  \citenamefont {Fields}, \citenamefont {Olive},\ and\ \citenamefont
  {Yeh}}]{Cyburt:2015mya}%
  \BibitemOpen
  \bibfield  {author} {\bibinfo {author} {\bibfnamefont {R.~H.}\ \bibnamefont
  {Cyburt}}, \bibinfo {author} {\bibfnamefont {B.~D.}\ \bibnamefont {Fields}},
  \bibinfo {author} {\bibfnamefont {K.~A.}\ \bibnamefont {Olive}}, \ and\
  \bibinfo {author} {\bibfnamefont {T.-H.}\ \bibnamefont {Yeh}},\ }\href
  {\doibase 10.1103/RevModPhys.88.015004} {\bibfield  {journal} {\bibinfo
  {journal} {Rev. Mod. Phys.}\ }\textbf {\bibinfo {volume} {88}},\ \bibinfo
  {pages} {015004} (\bibinfo {year} {2016})},\ \Eprint
  {http://arxiv.org/abs/1505.01076} {arXiv:1505.01076 [astro-ph.CO]}
  \BibitemShut {NoStop}%
\bibitem [{\citenamefont {Fradette}\ and\ \citenamefont
  {Pospelov}(2017)}]{Fradette:2017sdd}%
  \BibitemOpen
  \bibfield  {author} {\bibinfo {author} {\bibfnamefont {A.}~\bibnamefont
  {Fradette}}\ and\ \bibinfo {author} {\bibfnamefont {M.}~\bibnamefont
  {Pospelov}},\ }\href {\doibase 10.1103/PhysRevD.96.075033} {\bibfield
  {journal} {\bibinfo  {journal} {Phys. Rev.}\ }\textbf {\bibinfo {volume}
  {D96}},\ \bibinfo {pages} {075033} (\bibinfo {year} {2017})},\ \Eprint
  {http://arxiv.org/abs/1706.01920} {arXiv:1706.01920 [hep-ph]} \BibitemShut
  {NoStop}%
\bibitem [{\citenamefont {Wang}\ \emph {et~al.}(2014)\citenamefont {Wang},
  \citenamefont {Peter}, \citenamefont {Strigari}, \citenamefont {Zentner},
  \citenamefont {Arant}, \citenamefont {Garrison-Kimmel},\ and\ \citenamefont
  {Rocha}}]{Wang:2014ina}%
  \BibitemOpen
  \bibfield  {author} {\bibinfo {author} {\bibfnamefont {M.-Y.}\ \bibnamefont
  {Wang}}, \bibinfo {author} {\bibfnamefont {A.~H.~G.}\ \bibnamefont {Peter}},
  \bibinfo {author} {\bibfnamefont {L.~E.}\ \bibnamefont {Strigari}}, \bibinfo
  {author} {\bibfnamefont {A.~R.}\ \bibnamefont {Zentner}}, \bibinfo {author}
  {\bibfnamefont {B.}~\bibnamefont {Arant}}, \bibinfo {author} {\bibfnamefont
  {S.}~\bibnamefont {Garrison-Kimmel}}, \ and\ \bibinfo {author} {\bibfnamefont
  {M.}~\bibnamefont {Rocha}},\ }\href {\doibase 10.1093/mnras/stu1747}
  {\bibfield  {journal} {\bibinfo  {journal} {Mon. Not. Roy. Astron. Soc.}\
  }\textbf {\bibinfo {volume} {445}},\ \bibinfo {pages} {614} (\bibinfo {year}
  {2014})},\ \Eprint {http://arxiv.org/abs/1406.0527} {arXiv:1406.0527
  [astro-ph.CO]} \BibitemShut {NoStop}%
\bibitem [{\citenamefont {Klypin}\ \emph {et~al.}(1999)\citenamefont {Klypin},
  \citenamefont {Kravtsov}, \citenamefont {Valenzuela},\ and\ \citenamefont
  {Prada}}]{Klypin:1999uc}%
  \BibitemOpen
  \bibfield  {author} {\bibinfo {author} {\bibfnamefont {A.~A.}\ \bibnamefont
  {Klypin}}, \bibinfo {author} {\bibfnamefont {A.~V.}\ \bibnamefont
  {Kravtsov}}, \bibinfo {author} {\bibfnamefont {O.}~\bibnamefont
  {Valenzuela}}, \ and\ \bibinfo {author} {\bibfnamefont {F.}~\bibnamefont
  {Prada}},\ }\href {\doibase 10.1086/307643} {\bibfield  {journal} {\bibinfo
  {journal} {Astrophys. J.}\ }\textbf {\bibinfo {volume} {522}},\ \bibinfo
  {pages} {82} (\bibinfo {year} {1999})},\ \Eprint
  {http://arxiv.org/abs/astro-ph/9901240} {arXiv:astro-ph/9901240 [astro-ph]}
  \BibitemShut {NoStop}%
\bibitem [{\citenamefont {Boylan-Kolchin}\ \emph {et~al.}(2011)\citenamefont
  {Boylan-Kolchin}, \citenamefont {Bullock},\ and\ \citenamefont
  {Kaplinghat}}]{BoylanKolchin:2011de}%
  \BibitemOpen
  \bibfield  {author} {\bibinfo {author} {\bibfnamefont {M.}~\bibnamefont
  {Boylan-Kolchin}}, \bibinfo {author} {\bibfnamefont {J.~S.}\ \bibnamefont
  {Bullock}}, \ and\ \bibinfo {author} {\bibfnamefont {M.}~\bibnamefont
  {Kaplinghat}},\ }\href {\doibase 10.1111/j.1745-3933.2011.01074.x} {\bibfield
   {journal} {\bibinfo  {journal} {Mon. Not. Roy. Astron. Soc.}\ }\textbf
  {\bibinfo {volume} {415}},\ \bibinfo {pages} {L40} (\bibinfo {year}
  {2011})},\ \Eprint {http://arxiv.org/abs/1103.0007} {arXiv:1103.0007
  [astro-ph.CO]} \BibitemShut {NoStop}%
\bibitem [{\citenamefont {Jiang}\ and\ \citenamefont {van~den
  Bosch}(2015)}]{Jiang:2015vra}%
  \BibitemOpen
  \bibfield  {author} {\bibinfo {author} {\bibfnamefont {F.}~\bibnamefont
  {Jiang}}\ and\ \bibinfo {author} {\bibfnamefont {F.~C.}\ \bibnamefont
  {van~den Bosch}},\ }\href {\doibase 10.1093/mnras/stv1871} {\bibfield
  {journal} {\bibinfo  {journal} {Mon. Not. Roy. Astron. Soc.}\ }\textbf
  {\bibinfo {volume} {453}},\ \bibinfo {pages} {3575} (\bibinfo {year}
  {2015})},\ \Eprint {http://arxiv.org/abs/1508.02715} {arXiv:1508.02715
  [astro-ph.CO]} \BibitemShut {NoStop}%
\bibitem [{\citenamefont {Finkbeiner}\ and\ \citenamefont
  {Weiner}(2007)}]{Finkbeiner:2007kk}%
  \BibitemOpen
  \bibfield  {author} {\bibinfo {author} {\bibfnamefont {D.~P.}\ \bibnamefont
  {Finkbeiner}}\ and\ \bibinfo {author} {\bibfnamefont {N.}~\bibnamefont
  {Weiner}},\ }\href {\doibase 10.1103/PhysRevD.76.083519} {\bibfield
  {journal} {\bibinfo  {journal} {Phys. Rev.}\ }\textbf {\bibinfo {volume}
  {D76}},\ \bibinfo {pages} {083519} (\bibinfo {year} {2007})},\ \Eprint
  {http://arxiv.org/abs/astro-ph/0702587} {arXiv:astro-ph/0702587 [astro-ph]}
  \BibitemShut {NoStop}%
\bibitem [{\citenamefont {Huchra}\ \emph {et~al.}(2011)\citenamefont {Huchra},
  \citenamefont {Macri}, \citenamefont {Masters}, \citenamefont {Jarrett},
  \citenamefont {Berlind}, \citenamefont {Calkins}, \citenamefont {Crook},
  \citenamefont {Cutri}, \citenamefont {Erdogdu}, \citenamefont {Falco},
  \citenamefont {George}, \citenamefont {Hutcheson}, \citenamefont {Lahav},
  \citenamefont {Mader}, \citenamefont {Mink}, \citenamefont {Martimbeau},
  \citenamefont {Schneider}, \citenamefont {Skrutskie}, \citenamefont
  {Tokarz},\ and\ \citenamefont {Westover}}]{Huchra:2011aa}%
  \BibitemOpen
  \bibfield  {author} {\bibinfo {author} {\bibfnamefont {J.~P.}\ \bibnamefont
  {Huchra}}, \bibinfo {author} {\bibfnamefont {L.~M.}\ \bibnamefont {Macri}},
  \bibinfo {author} {\bibfnamefont {K.~L.}\ \bibnamefont {Masters}}, \bibinfo
  {author} {\bibfnamefont {T.~H.}\ \bibnamefont {Jarrett}}, \bibinfo {author}
  {\bibfnamefont {P.}~\bibnamefont {Berlind}}, \bibinfo {author} {\bibfnamefont
  {M.}~\bibnamefont {Calkins}}, \bibinfo {author} {\bibfnamefont {A.~C.}\
  \bibnamefont {Crook}}, \bibinfo {author} {\bibfnamefont {R.}~\bibnamefont
  {Cutri}}, \bibinfo {author} {\bibfnamefont {P.}~\bibnamefont {Erdogdu}},
  \bibinfo {author} {\bibfnamefont {E.}~\bibnamefont {Falco}}, \bibinfo
  {author} {\bibfnamefont {T.}~\bibnamefont {George}}, \bibinfo {author}
  {\bibfnamefont {C.~M.}\ \bibnamefont {Hutcheson}}, \bibinfo {author}
  {\bibfnamefont {O.}~\bibnamefont {Lahav}}, \bibinfo {author} {\bibfnamefont
  {J.}~\bibnamefont {Mader}}, \bibinfo {author} {\bibfnamefont {J.~D.}\
  \bibnamefont {Mink}}, \bibinfo {author} {\bibfnamefont {N.}~\bibnamefont
  {Martimbeau}}, \bibinfo {author} {\bibfnamefont {S.}~\bibnamefont
  {Schneider}}, \bibinfo {author} {\bibfnamefont {M.}~\bibnamefont
  {Skrutskie}}, \bibinfo {author} {\bibfnamefont {S.}~\bibnamefont {Tokarz}}, \
  and\ \bibinfo {author} {\bibfnamefont {M.}~\bibnamefont {Westover}},\ }\href
  {https://arxiv.org/abs/1108.0669} {\  (\bibinfo {year} {2011})},\ \Eprint
  {http://arxiv.org/abs/1108.0669} {1108.0669} \BibitemShut {NoStop}%
\bibitem [{\citenamefont {Kourkchi}\ and\ \citenamefont
  {Tully}(2017)}]{Kourkchi:2017aa}%
  \BibitemOpen
  \bibfield  {author} {\bibinfo {author} {\bibfnamefont {E.}~\bibnamefont
  {Kourkchi}}\ and\ \bibinfo {author} {\bibfnamefont {R.~B.}\ \bibnamefont
  {Tully}},\ }\href {https://arxiv.org/abs/1705.08068} {\  (\bibinfo {year}
  {2017})},\ \Eprint {http://arxiv.org/abs/1705.08068} {1705.08068}
  \BibitemShut {NoStop}%
\bibitem [{Note2()}]{Note2}%
  \BibitemOpen
  \bibinfo {note} {We thank Moira Gresham, Tim Lou, and Kathryn Zurek for
  emphasizing these essential points.}\BibitemShut {Stop}%
\bibitem [{\citenamefont {Baym}\ \emph {et~al.}(1971)\citenamefont {Baym},
  \citenamefont {Bethe},\ and\ \citenamefont {Pethick}}]{Baym:1971ax}%
  \BibitemOpen
  \bibfield  {author} {\bibinfo {author} {\bibfnamefont {G.}~\bibnamefont
  {Baym}}, \bibinfo {author} {\bibfnamefont {H.~A.}\ \bibnamefont {Bethe}}, \
  and\ \bibinfo {author} {\bibfnamefont {C.}~\bibnamefont {Pethick}},\ }\href
  {\doibase 10.1016/0375-9474(71)90281-8} {\bibfield  {journal} {\bibinfo
  {journal} {Nucl. Phys.}\ }\textbf {\bibinfo {volume} {A175}},\ \bibinfo
  {pages} {225} (\bibinfo {year} {1971})}\BibitemShut {NoStop}%
\bibitem [{\citenamefont {Mackie}\ and\ \citenamefont
  {Baym}(1977)}]{Mackie:1977nof}%
  \BibitemOpen
  \bibfield  {author} {\bibinfo {author} {\bibfnamefont {F.~D.}\ \bibnamefont
  {Mackie}}\ and\ \bibinfo {author} {\bibfnamefont {G.}~\bibnamefont {Baym}},\
  }\href {\doibase 10.1016/0375-9474(77)90256-1} {\bibfield  {journal}
  {\bibinfo  {journal} {Nucl. Phys.}\ }\textbf {\bibinfo {volume} {A285}},\
  \bibinfo {pages} {332} (\bibinfo {year} {1977})}\BibitemShut {NoStop}%
\bibitem [{\citenamefont {Detmold}\ \emph
  {et~al.}(2014{\natexlab{b}})\citenamefont {Detmold}, \citenamefont
  {McCullough},\ and\ \citenamefont {Pochinsky}}]{Detmold:2014kba}%
  \BibitemOpen
  \bibfield  {author} {\bibinfo {author} {\bibfnamefont {W.}~\bibnamefont
  {Detmold}}, \bibinfo {author} {\bibfnamefont {M.}~\bibnamefont {McCullough}},
  \ and\ \bibinfo {author} {\bibfnamefont {A.}~\bibnamefont {Pochinsky}},\
  }\href {\doibase 10.1103/PhysRevD.90.114506} {\bibfield  {journal} {\bibinfo
  {journal} {Phys. Rev.}\ }\textbf {\bibinfo {volume} {D90}},\ \bibinfo {pages}
  {114506} (\bibinfo {year} {2014}{\natexlab{b}})},\ \Eprint
  {http://arxiv.org/abs/1406.4116} {arXiv:1406.4116 [hep-lat]} \BibitemShut
  {NoStop}%
\bibitem [{Note3()}]{Note3}%
  \BibitemOpen
  \bibinfo {note} {We thank Moira Gresham, Tim Lou, and Kathryn Zurek for
  emphasizing these essential points.}\BibitemShut {Stop}%
\bibitem [{\citenamefont {Fan}\ \emph {et~al.}(2013{\natexlab{a}})\citenamefont
  {Fan}, \citenamefont {Katz}, \citenamefont {Randall},\ and\ \citenamefont
  {Reece}}]{Fan:2013yva}%
  \BibitemOpen
  \bibfield  {author} {\bibinfo {author} {\bibfnamefont {J.}~\bibnamefont
  {Fan}}, \bibinfo {author} {\bibfnamefont {A.}~\bibnamefont {Katz}}, \bibinfo
  {author} {\bibfnamefont {L.}~\bibnamefont {Randall}}, \ and\ \bibinfo
  {author} {\bibfnamefont {M.}~\bibnamefont {Reece}},\ }\href {\doibase
  10.1016/j.dark.2013.07.001} {\bibfield  {journal} {\bibinfo  {journal} {Phys.
  Dark Univ.}\ }\textbf {\bibinfo {volume} {2}},\ \bibinfo {pages} {139}
  (\bibinfo {year} {2013}{\natexlab{a}})},\ \Eprint
  {http://arxiv.org/abs/1303.1521} {arXiv:1303.1521 [astro-ph.CO]} \BibitemShut
  {NoStop}%
\bibitem [{\citenamefont {Fan}\ \emph {et~al.}(2013{\natexlab{b}})\citenamefont
  {Fan}, \citenamefont {Katz}, \citenamefont {Randall},\ and\ \citenamefont
  {Reece}}]{Fan:2013tia}%
  \BibitemOpen
  \bibfield  {author} {\bibinfo {author} {\bibfnamefont {J.}~\bibnamefont
  {Fan}}, \bibinfo {author} {\bibfnamefont {A.}~\bibnamefont {Katz}}, \bibinfo
  {author} {\bibfnamefont {L.}~\bibnamefont {Randall}}, \ and\ \bibinfo
  {author} {\bibfnamefont {M.}~\bibnamefont {Reece}},\ }\href {\doibase
  10.1103/PhysRevLett.110.211302} {\bibfield  {journal} {\bibinfo  {journal}
  {Phys. Rev. Lett.}\ }\textbf {\bibinfo {volume} {110}},\ \bibinfo {pages}
  {211302} (\bibinfo {year} {2013}{\natexlab{b}})},\ \Eprint
  {http://arxiv.org/abs/1303.3271} {arXiv:1303.3271 [hep-ph]} \BibitemShut
  {NoStop}%
\bibitem [{\citenamefont {McCullough}\ and\ \citenamefont
  {Randall}(2013)}]{McCullough:2013jma}%
  \BibitemOpen
  \bibfield  {author} {\bibinfo {author} {\bibfnamefont {M.}~\bibnamefont
  {McCullough}}\ and\ \bibinfo {author} {\bibfnamefont {L.}~\bibnamefont
  {Randall}},\ }\href {\doibase 10.1088/1475-7516/2013/10/058} {\bibfield
  {journal} {\bibinfo  {journal} {JCAP}\ }\textbf {\bibinfo {volume} {1310}},\
  \bibinfo {pages} {058} (\bibinfo {year} {2013})},\ \Eprint
  {http://arxiv.org/abs/1307.4095} {arXiv:1307.4095 [hep-ph]} \BibitemShut
  {NoStop}%
\bibitem [{\citenamefont {Boddy}\ \emph {et~al.}(2014)\citenamefont {Boddy},
  \citenamefont {Feng}, \citenamefont {Kaplinghat},\ and\ \citenamefont
  {Tait}}]{Boddy:2014yra}%
  \BibitemOpen
  \bibfield  {author} {\bibinfo {author} {\bibfnamefont {K.~K.}\ \bibnamefont
  {Boddy}}, \bibinfo {author} {\bibfnamefont {J.~L.}\ \bibnamefont {Feng}},
  \bibinfo {author} {\bibfnamefont {M.}~\bibnamefont {Kaplinghat}}, \ and\
  \bibinfo {author} {\bibfnamefont {T.~M.~P.}\ \bibnamefont {Tait}},\ }\href
  {\doibase 10.1103/PhysRevD.89.115017} {\bibfield  {journal} {\bibinfo
  {journal} {Phys. Rev.}\ }\textbf {\bibinfo {volume} {D89}},\ \bibinfo {pages}
  {115017} (\bibinfo {year} {2014})},\ \Eprint {http://arxiv.org/abs/1402.3629}
  {arXiv:1402.3629 [hep-ph]} \BibitemShut {NoStop}%
\bibitem [{\citenamefont {Boddy}\ \emph {et~al.}(2016)\citenamefont {Boddy},
  \citenamefont {Kaplinghat}, \citenamefont {Kwa},\ and\ \citenamefont
  {Peter}}]{Boddy:2016bbu}%
  \BibitemOpen
  \bibfield  {author} {\bibinfo {author} {\bibfnamefont {K.~K.}\ \bibnamefont
  {Boddy}}, \bibinfo {author} {\bibfnamefont {M.}~\bibnamefont {Kaplinghat}},
  \bibinfo {author} {\bibfnamefont {A.}~\bibnamefont {Kwa}}, \ and\ \bibinfo
  {author} {\bibfnamefont {A.~H.~G.}\ \bibnamefont {Peter}},\ }\href {\doibase
  10.1103/PhysRevD.94.123017} {\bibfield  {journal} {\bibinfo  {journal} {Phys.
  Rev.}\ }\textbf {\bibinfo {volume} {D94}},\ \bibinfo {pages} {123017}
  (\bibinfo {year} {2016})},\ \Eprint {http://arxiv.org/abs/1609.03592}
  {arXiv:1609.03592 [hep-ph]} \BibitemShut {NoStop}%
\bibitem [{\citenamefont {Agrawal}\ \emph {et~al.}(2017)\citenamefont
  {Agrawal}, \citenamefont {Cyr-Racine}, \citenamefont {Randall},\ and\
  \citenamefont {Scholtz}}]{Agrawal:2016quu}%
  \BibitemOpen
  \bibfield  {author} {\bibinfo {author} {\bibfnamefont {P.}~\bibnamefont
  {Agrawal}}, \bibinfo {author} {\bibfnamefont {F.-Y.}\ \bibnamefont
  {Cyr-Racine}}, \bibinfo {author} {\bibfnamefont {L.}~\bibnamefont {Randall}},
  \ and\ \bibinfo {author} {\bibfnamefont {J.}~\bibnamefont {Scholtz}},\ }\href
  {\doibase 10.1088/1475-7516/2017/05/022} {\bibfield  {journal} {\bibinfo
  {journal} {JCAP}\ }\textbf {\bibinfo {volume} {1705}},\ \bibinfo {pages}
  {022} (\bibinfo {year} {2017})},\ \Eprint {http://arxiv.org/abs/1610.04611}
  {arXiv:1610.04611 [hep-ph]} \BibitemShut {NoStop}%
\end{thebibliography}%

\end{document}